\documentclass[aip,amsmath,amssymb,reprint]{revtex4-1}

\usepackage{graphicx}
\usepackage{dcolumn}
\usepackage{bm}

\usepackage[utf8]{inputenc}
\usepackage[T1]{fontenc}
\usepackage{mathptmx}
\usepackage{etoolbox}

\usepackage{physics}
\usepackage{amsmath,dsfont}

\makeatletter
\def\@email#1#2{%
	\endgroup
	\patchcmd{\titleblock@produce}
	{\frontmatter@RRAPformat}
	{\frontmatter@RRAPformat{\produce@RRAP{*#1\href{mailto:#2}{#2}}}\frontmatter@RRAPformat}
	{}{}
}%
\makeatother

\begin{document}
	
\preprint{AIP/123-QED}

\title[High-Sensitivity Characterization of Atomic Layers using SHEL]{High-Sensitivity Characterization of Ultra-Thin Atomic Layers using Spin-Hall Effect of Light}

\author{Janmey J. Panda}

\author{Krishna R. Sahoo}

\author{Aparna Praturi}

\author{Ashique Lal}
\affiliation{Tata Institute of Fundamental Research - Hyderabad, Sy. No 36/P Serilingampally Mandal, Gopanpally Village, Hyderabad 500046, India}
\author{Nirmal K. Viswanathan}
\affiliation{School of Physics, University of Hyderabad, Hyderabad 500046, Telangana, India.}
\email{nirmalsp@uohyd.ac.in}

\author{Tharangattu N. Narayanan}
\email{tnn@tifrh.res.in}

\author{G. Rajalakshmi}
\email{raji@tifrh.res.in}
\affiliation{Tata Institute of Fundamental Research - Hyderabad, Sy. No 36/P Serilingampally Mandal, Gopanpally Village, Hyderabad 500046, India}

\begin{abstract}

The fast-emerging diverse applications using a variety of magnetic / non-magnetic heterostructure ultra-thin films warrant sensitive characterization of the electrical, optical and magnetic properties of the interface. As a practical alternate to the conventional magneto-optic Kerr effect (MOKE) method we propose and demonstrate spin-Hall effect of light (SHEL) based MOKE method with competitive sensitivity and scope for further improvement. The SHEL-MOKE technique is a versatile surface characterization tool for studying materials’ magnetic and dielectric ordering, which are extracted from the variations to the phase-polarization characteristics of a focused beam of light reflected at the interface, as a function of applied magnetic field. Using this technique, we measure magnetic field dependent complex Kerr angle and the coercivity in ultra-thin films of permalloy (Py) and at molybdenum disulphide (MoS$_2$) - permalloy (MSPy) hetero-structure interfaces. A comprehensive theoretical model and simulation data are provided to strengthen the potential of this simple non-invasive optical method. The theoretical model is subsequently applied to extract the optical conductivity of non-magnetic ultra-thin layers of MoS$_2$.

\end{abstract}

\maketitle
\section{Introduction}

Engineering spin-orbit coupling (SOC) and interfacial symmetry breaking in materials via magnetic - nonmagnetic interfacing are emerging trends in spintronics. The capability of SOC to enhance and tune the magnetic effects at the surface aids in the study of fundamental interactions at such surfaces and is also of significant interest in applications\cite{Dayen2020Two-dimensionalMagnetic-interfaces, Hellman2017Interface-inducedMagnetism,Zutic2004Spintronics:Applications}. Interfacing soft magnetic materials with atomically thin layers has been shown to improve the coercivity via SOC and strong short-range correlations between electronic spin and orbital degrees of freedom\cite{Hellman2017Interface-inducedMagnetism}. Understanding the modified magnetic properties at such interfaces is of paramount importance to adopt and tailor material for varied applications. In this context, the development of a simple but highly sensitive characterization tool for studying surface magnetic and dielectric properties holds importance.  

At present, vibrating sample magnetometry, magnetic force microscopy, and magneto-optic Kerr effect (MOKE) are widely used experimental methods for the investigation of magnetic properties of surface dominated atomic layers\cite{Bonilla2018StrongSubstrates,Kazakova2019FrontiersMicroscopy,Lan2020Magneto-opticApplications}. However, limitations arising due to large substrate contributions and measurement sensitivity are opening up opportunities for the development of alternate techniques \cite{deLimaBernardo2014SignalAmplification,Qiu2014DeterminationMeasurements,Du2020MeasurementMeasurement,Li2020MeasurementLight,You2021DetectionPointer,Li2019WeakLight}. Optical methods to characterize the magnetic properties of ultra-thin samples, through light-matter interaction, hold significant advantages including simple operation, non-contact probing, and high-sensitivity. One such promising method, proposed and demonstrated recently is the spin-Hall effect of light (SHEL)\cite{Hosten2008ObservationMeasurements,Ling2017RecentLight}, based on the weak measurement (WM) protocol of Aharanov et al.\cite{Aharonov1988How100}. The SHEL is a polarization-dependent transverse shift in the center of gravity of the optical beam trajectory arising due to spin-orbit interaction. By the appropriate selection of a pre and post-interaction state of polarization (SOP) pointer states of a paraxial beam of light, the WM method can significantly enhance the measurable signal beyond the system noise levels\cite{Hosten2008ObservationMeasurements,Yin2013PhotonicMetasurfaces,He2009MeasurementLight}. The SHEL is an archetypical example of the WM method, arising due to spin-orbit interaction (SOI) with underlying geometric phase\cite{Onoda2004HallLight}, and has revolutionized optical measurements, enabling extremely high precision measurements\cite{Chen2020PrecisionEffect,Zhou2018PhotonicMeasurements,Ling2017RecentLight}. Recently it was shown that SHEL measurements offer a new method to probe magnetic ordering in finite-sized 2D crystals \cite{Sahoo2020ProbingMaterials}. We expand on the basic idea to show that WM-based investigation is also suitable to measure the complex optical beam-shift to characterize and quantify the magnetization, coercivity, spin-orbit coupling strength, and the optical conductivity arising at the surface and interface of ultra-thin film samples. Although a few efforts have been made\cite{Du2020MeasurementMeasurement,Qiu2014DeterminationMeasurements,Li2019WeakLight,Li2020MeasurementLight,You2021DetectionPointer} in employing WM to study magnetic ordering in thin films, the concept was limited to certain techniques and orientations. A generalized theory was needed to understand and extract relevant material information from these experiments. We present here a generalized concept to understand the WM-based magneto-optical effects.

We demonstrate here our ability to tune the coercivity of ultra-thin films of permalloy (Py) via SOC by interfacing it with an atomically thin layer of molybdenum disulphide (MoS$_2$). The MoS$_2$ atomic layers are known to be high SOC material but lacking in magnetic ordering. The MoS$_2$, in proximity to the ferromagnetic Py film, modifies the magnetic ordering at the interface due to possible charge/spin transfer\cite{Yin2015FirstFe3O4/MoS2,Bansal2019ExtrinsicMoS2/Py}. The changes in interfacial magnetic ordering, in turn, change the SOP of the reflected light beam. Monitoring the changes in the SOP using the WM technique allows us an accurate and sensitive measurement of the interaction-dependent changes. This technique is referred to as the SHEL-MOKE method in this article. This SHEL-MOKE technique allows sensitive characterization of the enhancement in coercivity in polar and longitudinal MOKE configurations. The technique can be extended to different MOKE configurations too, leading to the possibilities of complete analyses of ultrathin magnetic films without using expensive magnetometers or other surface probing techniques. In addition, we also show that by careful designing of the SHEL-MOKE experiment, one can get information on the complex Kerr rotation – the Kerr angle and the Kerr ellipticity of the sample.We show both theoretically and experimentally how the SHEL-MOKE method can be employed to characterise ultra-thin films to extract not only their magnetic properties but their other optical properties too.

\begin{figure}[ht]
    \centering
    \includegraphics[width=\linewidth]{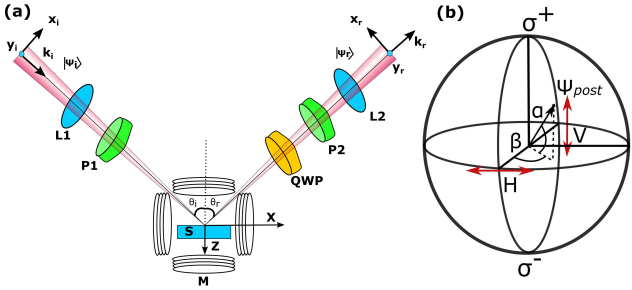}
    \caption{(a) Schematic of SHEL-MOKE experimental setup, where the sample (S) is placed in between the Helmholtz coil. Based on L-MOKE or P-MOKE measurement, the magnetic field direction can be changed to parallel or perpendicular to the sample surface. (b)  Bloch sphere representation of polarization state.}
    \label{fig:sch}
\end{figure}

\section{Theory of spin-Hall effect of light in magnetic films}

When a paraxial beam of light is reflected or transmitted at an air-dielectric interface (including the magnetic thin film interface presented here), the phase-polarization characteristics in the beam cross-section get modified significantly. These changes depend on the SOP of the incident beam, angle of incidence, interface characteristics, and on the externally applied field \cite{Ling2017RecentLight}. Monitoring the phase-polarization changes in the interacted beam via the weak measurement method provides us with invaluable information about the interface characteristics up to a level that is much below the sensitivity of most existing methods \cite{Hosten2008ObservationMeasurements,Yin2013PhotonicMetasurfaces,He2009MeasurementLight}. To understand in detail the interface characteristics and changes arising in it due to the light-matter interaction, we consider a monochromatic linearly-polarized Gaussian beam of frequency $\omega$, incident on the interface between a lossless non-magnetic and magnetic media. The dielectric properties of magnetic thin film are effectively described by the tensor \textbf{$\varepsilon$} given by\cite{Qiu2000SurfaceEffect,Yang1998CombinedMagnetism}:
\begin{equation}
    \varepsilon= \epsilon \begin{bmatrix}
     1 && -iQ_{z} && iQ_{y}\\
    iQ_{z} && 1 && -iQ_{x}\\
    -iQ_{y} && iQ_{x} && 1 
    \end{bmatrix}
\end{equation}

Here, $\epsilon$ is the dielectric constant of the material, and $Q_{x,y,z}$ represents the magneto-optic Voigt constants of the magnetic material in x,y, and z directions respectively. In the right-handed Cartesian coordinate system (X, Y, Z), the (X, Y) plane forms the interface at Z=0, with the YZ plane taken as the plane of incidence Fig. \ref{fig:sch}(a). The light beam is described in a coordinate system defined by ($x^{a},y^{a},z^{a}$), where superscript a= {(i), (r)} denotes the incident and reflected light respectively. In this coordinate system $y^{a}$-axis coincides with the Y-axis and $z^{a}$ is directed along the wave vector of the beam’s central plane corresponding to k$^{a}_{\circ}$. A linearly polarized Gaussian beam propagating in this coordinate system can be expressed in momentum space representation as:

\begin{equation}
     \ket{\Psi(\textbf{k})}^{a} = \ket{\psi(\textbf{k})}^{a} \phi^{a}_{k}
\end{equation}
with
\begin{eqnarray}\label{3}
\ket{\psi}^{a} & = & E_{H}^{a}\ket{H}^{a} + E_{V}^{a}\ket{V}^{a}\\ \nonumber
\phi^{a}_{k} & = & \frac{\omega_{0}}{\sqrt{2\pi}}\exp{-(\frac{iz}{2k_{o}}+\frac{\omega^{2}_{0}}{4})((k^{a}_{x})^{2}+(k^{a}_{y})^{2})}
\end{eqnarray}

Here, $\ket{\psi}^{a}$ refers to the polarization state while  $\phi^{a}_{k}$ describes the Gaussian profile of the incident (reflected) beam with $\omega_{0}$ and z representing the beam waist and propagation distance respectively. Since, we mostly work in k-space,  explicit mention of  \textbf{k} is dropped in the function representations. The coordinate space functions are represented with a tilde. $k^{a}_{x}$ and $k^{a}_{y}$ represent the transverse spread of wave vectors corresponding to $x^{a}$ and $y^{a}$ coordinate axis. $\ket{H}^{a}(\ket{V}^{a})$ represents the horizontal (vertical) polarization state of light directed along $(\hat{x}^{a}-\frac{k_{x}^{a}}{k_{\circ}^{a}}\hat{z}^{a})$ and $(\hat{y}^{a}-\frac{k_{y}^{a}}{k_{\circ}^{a}}\hat{z}^{a})$ axis\cite{BliokhPolarizationPacket,Hosten2008ObservationMeasurements}, which corresponds to transverse magnetic mode (transverse electric mode) of light and $E^{a}_{H}(E^{a}_{V})$ is the corresponding electric field amplitude.
A Gaussian beam with spherical wavefront can be described by an infinite set of plane waves with wave vectors $\textbf{k}^{a} =k^{a}_{x} \hat{x}^{a}+k^{a}_{y} \hat{y}^{a} + k^{a}_{\circ} \hat{z}^{a}$  enclosing the central wave vector $k^{a}_{\circ} \hat{z}^{a}$ \cite{BliokhPolarizationPacket}. Under the paraxial approximation, the spread of the beam perpendicular to the central propagation direction given by  $k^{a}_{x}, k^{a}_{y}$, are in general much smaller than the central wave vector $k^{a}_{\circ}=k_{\circ}=\frac{2\pi}{\lambda}$. Fresnel reflection coefficients relate the  incident and reflected plane wave vector and describe the action of the interface. Thus, the reflection of Gaussian beam can be effectively modeled by using Fresnel reflection coefficients acting on individual incident plane wave vectors. The electric field components of  H and V  in the incident and reflected Gaussian beam can be related using reflection matrix R as $\big(\begin{smallmatrix}E^{r}_{H}\\E^{r}_{V}\end{smallmatrix}\big)=R\big(\begin{smallmatrix}E^{i}_{H}\\E^{i}_{V}\end{smallmatrix}\big)$, where R can be obtained using the approach as mentioned in the references \cite{Luo2011EnhancingNanostructures,Bliokh2013GoosHanchenOverview} by relating incident and reflected Gaussian beam using Fresnel reflection matrix for a magnetic medium. R is a $2\cross2$ matrix given by, 
\begin{equation} \label{4}
    R=\begin{bmatrix}r^{pp}-\frac{k^{r}_{y}(r^{ps}-r^{sp})\cot{\theta_{i}}}{k_{\circ}}&&r^{ps}+\frac{k^{r}_{y}(r^{pp}+r^{ss})\cot{\theta_{i}}}{k_{\circ}}\\r^{sp}-\frac{k^{r}_{y}(r^{pp}+r^{ss})\cot{\theta_{i}}}{k_{\circ}}&&r^{ss}-\frac{k^{r}_{y}(r^{ps}-r^{sp})\cot{\theta_{i}}}{k_{\circ}}
    \end{bmatrix}.
\end{equation}
where, $r^{pp}, r^{ss}, r^{sp}, r^{ps}$ are the Fresnel reflection coefficient of the sample [S3-S6]. These coefficients are, in general, derived by applying boundary conditions on the electric and magnetic field components of the electromagnetic wave at the material interface. For reflection from magnetic material interface one can easily find these components by enforcing the boundary conditions as specified by Zak et al. \cite{Zak1990UniversalMagneto-optics}. Considering the fact that Gaussian beam also has a small spread of wave vectors in the plane of incidence, the Fresnel coefficients can be Taylor expanded to first-order approximation as\cite{BliokhPolarizationPacket,Pan2013ImpactAngle}: $r^{ss}=r_{\circ}^{ss}-\frac{k^{r}_{x}}{k_{\circ}}\frac{\partial{r^{ss}}}{\partial{\theta_i}}$, $r^{pp}=r_{\circ}^{pp}-\frac{k^{r}_{x}}{k_{\circ}}\frac{\partial{r^{pp}}}{\partial{\theta_{i}}}$, $r^{sp}=r_{\circ}^{sp}-\frac{k^{r}_{x}}{k_{\circ}}\frac{\partial{r^{sp}}}{\partial{\theta_{i}}}$ and $r^{ps}=r_{\circ}^{ps}-\frac{k^{r}_{x}}{k_{\circ}}\frac{\partial{r^{ps}}}{\partial{\theta_{i}}}$. Relating the incident and reflected wavevectors using $\ket{\psi}^{r}=R\ket{\psi}^{i}$, the reflection matrix R in equation (4) can now be rewritten as,
\begin{equation} \label{5}
    R=\left( \begin{smallmatrix}r^{pp}_{\circ}-\frac{k^{r}_{x}}{k_{\circ}}\frac{\partial{r^{pp}}}{\partial{\theta_{i}}}-\frac{k^{r}_{y}(r^{ps}_{\circ}-r^{sp}_{\circ})\cot{\theta_{i}}}{k_{\circ}}&&r^{ps}_{\circ}-\frac{k^{r}_{x}}{k_{\circ}}\frac{\partial{r^{ps}}}{\partial{\theta_{i}}}+\frac{k^{r}_{y}(r^{pp}_{\circ}+r^{ss}_{\circ})\cot{\theta_{i}}}{k_{\circ}}\\r^{sp}_{\circ}-\frac{k^{r}_{x}}{k_{\circ}}\frac{\partial{r^{sp}}}{\partial{\theta_{i}}}-\frac{k^{r}_{y}(r^{pp}_{\circ}+r^{ss}_{\circ})\cot{\theta_{i}}}{k_{\circ}}&&r^{ss}_{\circ}-\frac{k^{r}_{x}}{k_{\circ}}\frac{\partial{r^{ss}}}{\partial{\theta_{i}}}-\frac{k^{r}_{y}(r^{ps}_{\circ}-r^{sp}_{\circ})\cot{\theta_{i}}}{k_{\circ}}
    \end{smallmatrix}\right).
\end{equation}

In the spin basis, the horizontal and vertical polarization of light can be written as $\ket{H}^{a}=\frac{\ket{+}^{a}+\ket{-}^{a}}{\sqrt{2}}$ and $\ket{V}^{a}=\frac{i(\ket{-}^{a}-\ket{+}^{a})}{\sqrt{2}}$, where $\ket{+}^{a}$ and $\ket{-}^{a}$ denotes the left and right circular polarization state respectively. The components of circular polarization of reflected light can be related to incident light in linear polarization as $\big(\begin{smallmatrix}E^{r}_{+}\\E^{r}_{-}\end{smallmatrix}\big)=\frac{R^{c}}{\sqrt{2}} \big(\begin{smallmatrix}E^{i}_{H}\\E^{i}_{V}\end{smallmatrix}\big)$, where, $R^{c}$ takes care of the transformation from linear to circular polarization basis of the incident beam as well as the reflection:
\begin{widetext}
\begin{equation}\label{6}
    R^{c}= \left(
    \begin{matrix} r_{\circ}^{pp}[1+ik^{r}_{y}\delta_{H}-k^{r}_{x}\zeta_{H}-i\theta_{H}[1-k^{r}_{x}\zeta_{H}^{\prime}+ik^{r}_{y} \delta_{H}^{\prime}]] && -ir_{\circ}^{ss}[1+ik^{r}_{y}\delta_{V}-k^{r}_{x}\zeta_{V}+i\theta_{V}[1-k^{r}_{x}\zeta_{V}^{\prime}+ik^{r}_{y} \delta^{\prime}_{V}]]\\ r_{\circ}^{pp}[1-ik^{r}_{y}\delta_{H}-k^{r}_{x}\zeta_{H}+i\theta_{H}[1-k^{r}_{x}\zeta_{H}^{\prime}-ik^{r}_{y} \delta_{H}^{\prime}]] && ir_{\circ}^{ss}[1-ik^{r}_{y}\delta_{V}-k^{r}_{x}\zeta_{V}-i\theta_{V}[1-k^{r}_{x}\zeta_{V}^{\prime}-ik^{r}_{y} \delta^{\prime}_{V}]]
    \end{matrix}
	\right).
\end{equation}
\end{widetext}
Here, $\delta_{H}=\frac{\cot{\theta_{i}}}{k_{\circ}}(1+r_{\circ}^{ss}/r_{\circ}^{pp})$, $\delta^{\prime}_{H}=\frac{\cot{\theta_{i}}}{k_{\circ}}(1-r_{\circ}^{ps}/r_{\circ}^{sp})$, $\delta_{V}=\frac{\cot{\theta_{i}}}{k_{\circ}}(1+r_{\circ}^{pp}/r_{\circ}^{ss})$, $\delta^{\prime}_{V}=\frac{\cot{\theta_{i}}}{k_{\circ}}(1-r_{\circ}^{sp}/r_{\circ}^{ps})$, $\zeta_{H}=\frac{1}{k_{\circ}r^{pp}_{\circ}}\frac{\partial{r^{pp}}}{\partial{\theta_{i}}}$, $\zeta_{H}^{\prime}=\frac{1}{k_{\circ}r^{sp}_{\circ}}\frac{\partial{r^{sp}}}{\partial{\theta_{i}}}$, $\zeta_{V}=\frac{1}{k_{\circ}r^{ss}_{\circ}}\frac{\partial{r^{ss}}}{\partial{\theta_{i}}}$, $\zeta_{V}^{\prime}=\frac{1}{k_{\circ}r^{ps}_{\circ}}\frac{\partial{r^{ps}}}{\partial{\theta_{i}}}$ $\theta_{H}=r^{sp}_{\circ}/r^{pp}_{\circ}$, and $\theta_{V}=r^{ps}_{\circ}/r^{ss}_{\circ}$\cite{Zak1990UniversalMagneto-optics,Qiu2014DeterminationMeasurements}. In equation \eqref{6}, $(1\pm ik^{r}_{y}\delta_{H,V)}$ represents the spin orbit interaction~(SOI) of light due to the diagonal component of Fresnel reflection coefficients. This SOI term is the main contributor to the initial spin dependent shift in SHEL. On the other hand, $(1\pm ik^{r}_{y}\delta^{\prime}_{H,V})$ represent the SOI of light due to off-diagonal Fresnel reflection coefficients. This term along with, $\theta_H$ and  $\theta_V$, represents the optical activity of the material and mostly acts as perturbation to the initial spin dependent shift. 

In general, for magnetic materials displaying the Kerr effect, the complex Kerr angle for H and V polarized incident light are represented by  $\theta_{H}$ and $\theta_{V}$,  respectively. The real ($\theta_{r}$) and imaginary ($\theta_{e}$) part of the complex Kerr angle represents the Kerr rotation of polarization and Kerr ellipticity change due to change in sample magnetization. Using equations S5 and S6, it is apparant that $\delta^{\prime}_{(H/V)}$ vanishes for polar and transverse magnetization direction whereas it has finite contribution in spin separation for longitudinal magnetization direction.
 
The SHEL shift of the reflected beam is obtained by calculating the expectation value of the transverse position operator. However, in real scenarios the actual shift is usually very small, approximately a fraction of a wavelength, which cannot be detected using conventional photodetectors or cameras. To efficiently detect the spin-dependent shift and the effect of sample magnetization on this shift, the principles of weak measurements are used to obtain an amplified spin-dependent shift. The amplified shift can be easily detected using a regular imaging camera. The ideal way of using the weak measurement to amplify spin-dependent shift lies in using H or V linear polarization as the initial state and projecting to the nearly orthogonal polarization state for detection. The incident polarization state $\ket{\psi}^{i}$ after reflection evolves to $\ket{\psi_{evl}}_{(H/V)}= (E^{r}_{+}\ket{+}^{r} + E^{r}_{-}\ket{-}^{r})_{(H/V)}$ in the circular basis. Here the subscript H and V represents the initial incident light polarization in H or V direction i.e., $E^{i}_{H}=1,E^{i}_{V}=0$ for H polarization and $E^{i}_{H}=0,E^{i}_{V}=1$ for V polarization. This evolved state is then observed using an analyzer in nearly orthogonal polarization configuration $\ket{\psi_{post}}$. Using the expressions for $E^{r}_{+}$ and $E^{r}_{-}$, the evolved light can be written as:

 \begin{equation}
\begin{aligned}\label{7}
\ket{\psi_{evl}}_{(H)} = &\frac{r^{pp}_{\circ}}{\sqrt{2}}\Big[\big(1+ik^{r}_{y}\delta_{H}-k^{r}_{x}\zeta_{H}-i\theta_{H}[1-k^{r}_{x}\zeta_{H}^{\prime}+ik^{r}_{y} \delta_{H}^{\prime}]\big)\ket{+}^{r} \\
&+\big(1-ik^{r}_{y}\delta_{H}-k^{r}_{x}\zeta_{H}+i\theta_{H}[1-k^{r}_{x}\zeta_{H}^{\prime}-ik^{r}_{y} \delta_{H}^{\prime}]\big)\ket{-}^{r} \Big]\\
\sim  & r^{pp}_{\circ}\left[\hat{\mathds{1}}+ik^{r}_{y}\delta_{H}\hat{A}-k^{r}_{x}\zeta_{H}\hat{\mathds{1}} \right. \\
&-i\theta_{H}(\hat{A}-k^{r}_{x}\zeta_{H}^{\prime}\hat{A}+ik^{r}_{y}\delta_{H}^{\prime}\hat{\mathds{1}})\left. \right]\ket{\psi_{pre}}_{(H)}.
\end{aligned}
\end{equation}
\begin{equation}\label{8}
\begin{aligned}
\ket{\psi_{evl}}_{(V)} = &\frac{-ir^{ss}_{\circ}}{\sqrt{2}}\Big[\big(1+ik^{r}_{y}\delta_{V}-k^{r}_{x}\zeta_{V}+i\theta_{V}[1-k^{r}_{x}\zeta_{V}^{\prime}+ik^{r}_{y} \delta^{\prime}_{V}]\big)\ket{+}^{r}\\
&-\big(1-ik^{r}_{y}\delta_{V}-k^{r}_{x}\zeta_{V}-i\theta_{V}[1-k^{r}_{x}\zeta_{V}^{\prime}-ik^{r}_{y} \delta^{\prime}_{V}]\big)\ket{-}^{r}\Big]\\
\sim  & r^{ss}_{\circ}\left[\hat{\mathds{1}}+ik^{r}_{y}\delta_{V}\hat{A}-k^{r}_{x}\zeta_{V}\hat{\mathds{1}} \right. \\
&+i\theta_{V}(\hat{A}-k^{r}_{x}\zeta_{V}^{\prime}\hat{A}+ik^{r}_{y}\delta_{V}^{\prime}\hat{\mathds{1}})\left. \right]\ket{\psi_{pre}}_{(V)}.
\end{aligned}
\end{equation}
Here,  $\ket{\psi_{pre}}_{(H)}=\frac{1}{\sqrt{2}}(\ket{+}+\ket{-})$,  $\ket{\psi_{pre}}_{(V)}=\frac{i}{\sqrt{2}}(\ket{-}-\ket{+})$, and $\hat{A} =\big(\begin{smallmatrix}1&&0\\0&&-1\end{smallmatrix}\big)$. To take advantage of the weak value amplification, post-selection is performed on the evolved state of light. A quarter wave plate (QWP) and an analyzer that selects orthogonal polarization to the initial state perform the post selection. Their action can be represented in the circular basis as \cite{Zhou2014OptimalEffect}:
\begin{equation}\label{9}
\begin{split}
    \ket{\psi_{post}}_{(H)}&= i\big[(e^{i\beta}\cos{(\frac{\pi}{4}+\alpha)}\ket{-}- e^{-i\beta}\sin{(\frac{\pi}{4}+\alpha)}\ket{+}\big].\\
    \ket{\psi_{post}}_{(V)}&= e^{-i\beta}\cos{(\frac{\pi}{4}+\alpha)}\ket{+}+ e^{i\beta}\sin{(\frac{\pi}{4}+\alpha)}\ket{-}.
\end{split}
\end{equation}
where $\alpha$ is small deviation of the QWP optic axis from the initial polarization direction and $\beta$ is the  small angular deviation from the perfect orthogonal position of the analyser optic axis. There actions on the light beam  are represented on the Bloch sphere  in Fig.\ref{fig:sch}(b). Hence, the final state of reflected light is represented by the projection of $\ket{\psi_{evl}}_{(H/V)}$ on the post selected polarization state $\ket{\psi_{post}}_{(H/V)}$. Mathematically, the weak measurement on evolved state of light i.e., $\ket{\psi_{final}}_{(H)} = \ket{\psi_{post}}\bra{\psi_{post}}\ket{\psi_{evl}}_{(H)}$ and  $\ket{\psi_{final}}_{(V)} = \ket{\psi_{post}}\bra{\psi_{post}}\ket{\psi_{evl}}_{(V)}$ can be written in momentum space as shown in equation \eqref{10}. In the coordinate space, it is given by the angular spectrum representation as in equation \eqref{11}:
\begin{equation}\label{10}
\begin{aligned}
\ket{\psi_{final}}_{(H)} = & \ \gamma_{(H)}(1+ik^{r}_{y}\delta_{H}A_{(H)}^{W}-k^{r}_{x}\zeta_{(H)}\\
-i\theta_{(H)}(&A_{(H)}^{W}-k^{r}_{x}\zeta_{H}^{\prime}A_{(H)}^{W}-ik^{r}_{y}\delta^{\prime}_{H}))\ket{\psi_{post}}\bra{\psi_{post}}\ket{\psi_{pre}}_{(H)}
\end{aligned}
\end{equation}
\begin{equation}
\begin{aligned}
\ket{\psi_{final}}_{(V)} = & \  \gamma_{(V)}(1+ik^{r}_{y}\delta_{V}A_{(V)}^{W}-k^{r}_{x}\zeta_{(V)}\\
+i\theta_{(V)}(&A_{(V)}^{W}-k^{r}_{x}\zeta_{V}^{\prime}A_{(V)}^{W}-ik^{r}_{y}\delta^{\prime}_{V}))\ket{\psi_{post}}\bra{\psi_{post}}\ket{\psi_{pre}}_{(V)}
\end{aligned}
\end{equation}
\begin{equation}\label{11}
	\begin{aligned}
    & \ket{\tilde{\Psi}_{final}}_{(H/V)}  =  \\ 
    & \iint_{-\infty}^{\infty}\exp{i(k^{r}_{x}x^{r}+k^{r}_{y}y^{r}+k_{\circ}z^{r})}\ket{\psi_{final}}_{(H/V)}\phi^{r}_{k}\,dk^{r}_{x},
\,dk^{r}_{y}.
\end{aligned}
\end{equation}
Here $\gamma_{(H/V)}=r^{pp}_{\circ}\ \textrm{or}\ r^{ss}_{\circ}$ for  H or V pre-selected state respectively and the analytical expression for corresponding weak value amplification factor $A_{(H/V)}^{W}$ is given by:
\begin{equation}\label{12}
    A_{(H/V)}^{W}=\frac{\bra{\psi_{post}}\hat{A}\ket{\psi_{pre}}_{(H/V)}}{\bra{\psi_{post}}\ket{\psi_{pre}}_{(H/V)}} \sim \frac{\pm \sin{2\alpha}+i\sin{2\beta}\cos{2\alpha}}{2\abs{\bra{\psi_{post}}\ket{\psi_{pre}}_{(H/V)}}^{2}}.
\end{equation}

From the above relation, it is obvious that a large weak value can be obtained when the pre and post-selection states are nearly orthogonal. Using equation \eqref{11} and equation \eqref{12} we can model the amplified spin shift by calculating the expectation value of the transverse position operator using,
 
\begin{equation}\label{13}
\Delta^{mod}_{(H/V)}=\frac{\bra{\tilde{\Psi}_{final}}y\ket{\tilde{\Psi}_{final}}_{(H/V)}}{\bra{\tilde{\Psi}_{final}}\ket{\tilde{\Psi}_{final}}_{(H/V)}}.
\end{equation}
The dependence of modified shift on complex Kerr angle is obtained by solving equation \eqref{13} and is provided in equation S1 and equation S2 in supporting information [SI] for initial H and V polarized light respectively.
Considering the demagnifying factor (F) introduced by the use of a second lens in order to contain the image within the camera area, the observed amplified spin shift is obtained by multiplying F to equation \eqref{13}. In the absence of magnetic field dependence and for samples in which the refractive index is purely real, the modified shift is reduced to:
{\small 
	\begin{equation} \label{16}
    \Delta^{mod}_{H/V}=F
\frac{2\gamma_{H/V}(r^{pp}_{\circ}+r^{ss}_{\circ})z\cot{\beta}\cot{\theta_i}}{(k_{\circ} \omega_{0}\gamma_{H/V})^2 +k_{\circ}^{2}(\gamma_{H/V})^2\zeta^{2}_{H/V}+(r^{pp}_{\circ}+r^{ss}_{\circ})^{2}\cot^{2}{\beta}\cot^{2}{\theta_{i}}}.
\end{equation}
}
\begin{figure}[t]
    \centering
    \includegraphics[width=\linewidth]{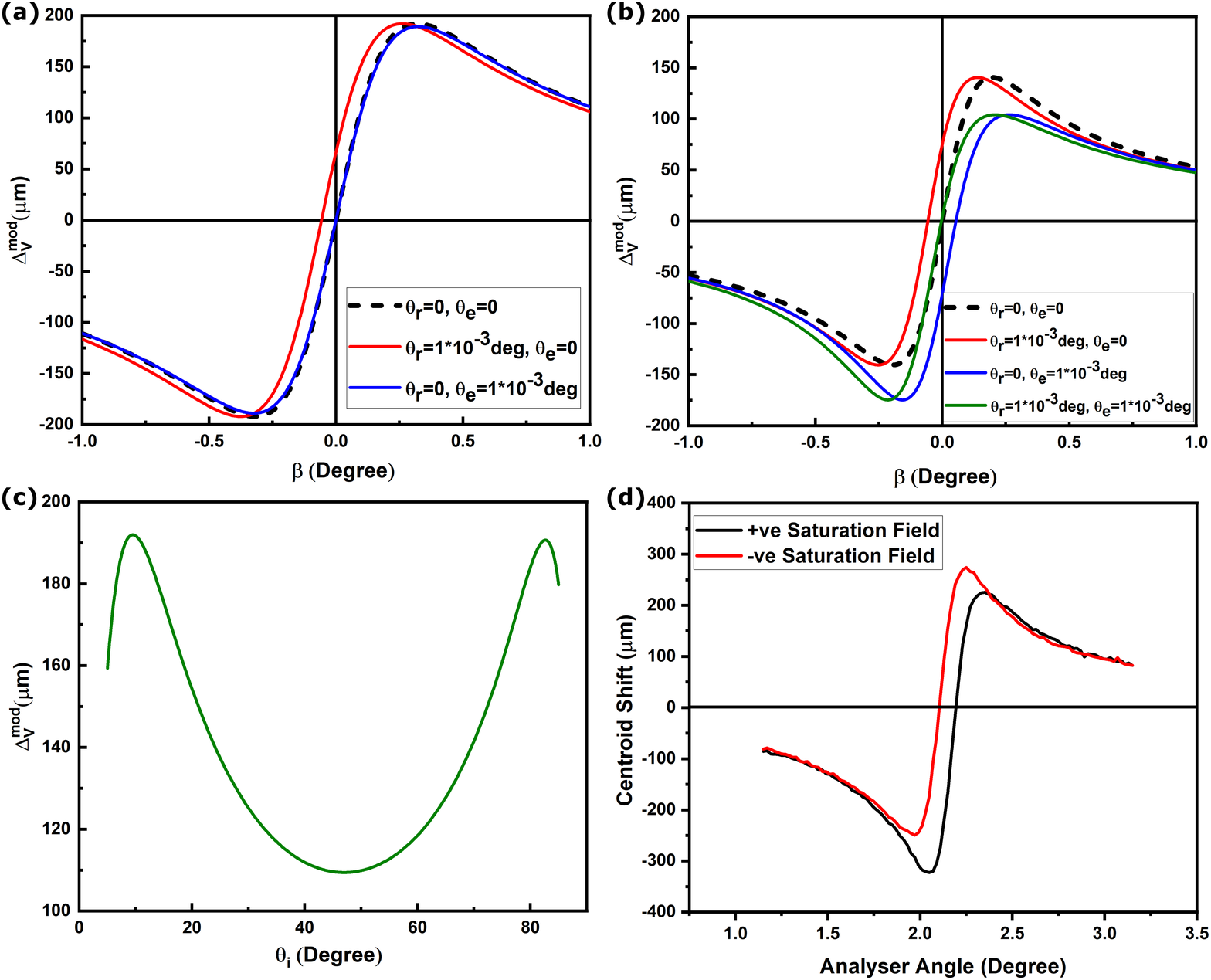}
    \caption{Plots of modified shift for material with  (a) the real value of the refractive index, and (b) complex value of the refractive index at different conditions of Kerr rotation and ellipticity as a function of change in post-selection angle ($\beta$), (c) the real value of the refractive index as a function of angle of incidence at $\beta=0.1$ deg, (d) experimental SHEL plot for permalloy of 20 nm thickness (Py20) at positive and negative saturation magnetizations acquired using 632.8 nm laser source.}
    \label{fig:Model}
\end{figure}

For a V-polarized incident light, Fig.\ref{fig:Model}(a) shows the plot of the expected modified shift as a function of analyzer angle away from orthogonal position ($\beta$) for a system with real refractive index (n=1.45), in the absence of magnetic field. When the system does not have any remnant Kerr rotation or ellipticity in zero field, the shift is as given by the dashed black line in the plot. If we introduce a zero field polarization rotation $\theta_r = 1\times 10^{-3}$~deg , the curve moves along the $\beta$ axis without any change in slope (red curve, Fig.\ref{fig:Model}(a)). The presence of a small rotation leads to a large change in the analyzer angle for the crossed position, $\beta_0$. The presence of zero field ellipticity on the other hand changes the slope of the curve without any change in the zero shift position (blue curve, Fig.\ref{fig:Model}(a)). The small change in slope is accompanied by a reduction in the maximum observed shift. However, the behavior in the case of materials with a complex refractive index (n=1.45 + i 1) is quite different for the same condition, as shown in Fig.\ref{fig:Model}(b). A zero field polarization rotation causes change in $\beta_0$ corresponding to zero shift position (red curve, Fig.\ref{fig:Model}(b)) similar to the earlier case, but zero field ellipticity gives rise to change in $\beta_0$ as well as asymmetry about the $\beta$ axis(blue curve, Fig.\ref{fig:Model}(b)). The maximum values of positive and negative shifts are not the same, but there is no change in the slope of the curve. In the presence of equal zero field rotation and ellipticity, the change along the $\beta$ axis is compensated, showing that contributions of both to the centroid shift are in opposite directions. Fig.\ref{fig:Model}(c) shows the dependence of centroid shift on the angle of incidence for materials with a real value of the refractive index at $\beta=0.1$ deg. For confirmation, we performed SHEL measurement on Py samples which have a complex refractive index at two different saturation magnetization fields. The plots are shown in Fig. \ref{fig:Model}(d). The curves for the positive and negative saturation magnetizations have the same slope, but the analyzer angle for zero shift changes due to the modification in the SOP of light, depicting the presence of both rotation and ellipticity. To study the material's magnetic properties, SHEL-MOKE has been performed and the results are compared with MOKE. The details of experimental methods and the results are discussed in the following sections.  

\section{Experimental methods}
To determine the improved measurement capabilities of the SHEL-MOKE and to compare its performance with the standard MOKE setup, permalloy thin films (Py, Ni$_{80}$Fe$_{20}$), a well studied ferromagnetic material, and molybdenum disulphide (MoS$_2$)/Py samples (MSPy) are prepared and used to study the effects arising due to spin-orbit coupling (SOC). We prepared five different samples of Permalloy (Py) thin films. Among them, one sample contains molybdenum disulphide (MoS$_2$)/Py interface (MSPy). The five samples are refereed as Py10, Py20, Py30, Py45, and MSPy45, where the number at the end represents Py thickness in nanometer (nm). The details of the sample preparation are given below:

\subsection{Permalloy thin film deposition}\label{2.1}

Thin films of Py are deposited by thermal evaporation of Py pellets on Si/SiO$_{2}$ substrate (wafer, 300 nm SiO$_{2}$ on a silicon (Si) wafer) and SiO$_{2}$/MoS$_2$ monolayers. A slow deposition rate of (0.5-0.6 \AA \ /sec) is maintained by controlling the filament current, to obtain a uniform thickness of Py thin films. The thickness of Py deposited was monitored using a quartz crystal-based thickness monitor. The chamber pressure was maintained at $5.5 \times 10^{-6}$ mbar during the deposition. All samples described here were deposited under the same conditions of temperature and pressure. The deposited thin films are characterized using a scanning electron microscope (SEM) imaging and energy-dispersive X-ray spectroscopy (EDS) mapping techniques confirming the absence of impurities Fig.S1 [SI]. The SEM-EDS mapping confirms the formation of Py film with a uniform elemental distribution throughout the sample.  

\subsection{Synthesis of MoS$_2$}
 
A custom-built chemical vapor deposition (CVD) apparatus was used for the synthesis of MoS$_2$ monolayers. The growth is performed using a two-zone furnace under atmospheric pressure with precursor (MoO$_3$) placed at 700 $^{\circ}$C and sulfur at 200 $^{\circ}$C. A constant flow rate of nitrogen gas was maintained at 185 sccm inside the furnace. Typically, a Si/SiO$_{2}$(300 nm) substrate is placed on top of the precursor where the MoS$_2$ monolayers will form\cite{Sharma2019OnGeneration,Mallik2021Salt-assistedTransistor}. The MoS$_2$ sample formed by the above process was characterized using micro Raman spectroscopy as shown in Fig.S2 [SI]. The difference between Raman vibration modes of E$_{2g}$ and A$_{1g}$  confirms the formation of monolayer MoS$_2$ 2H-phase on top of Si/SiO$_2$ substrate. A ferromagnet MoS$_2$ interface is then made by depositing Py thin film using methods described in section 3.1. Raman signal acquired on this MSPy sample shows intact phase and quality of underlying  MoS$_2$ monolayers. Hence the thermal evaporation of Py did not seem to be affected the quality and phase of the CVD grown MoS$_2$..

\subsection{MOKE} 

For the initial characterization of magneto-optic properties of samples, room temperature magneto-optic Kerr effect (MOKE) measurement is performed using a homemade MOKE setup. Schematic of the setup is shown in, Fig.S3 [SI]. The setup uses a 632.8 nm linearly polarized He-Ne laser source with an output power of 10 mW to carry out the Kerr optical rotation measurements in both longitudinal (L-) and polar (P-) geometries. A Helmholtz coil, capable of sweeping fields up to 100 Gauss, is used for changing the sample magnetization. The polarization rotation introduced by the sample due to an applied magnetic field is observed using an analyzer and a photodetector. The signal-to-noise ratio of the measured optical rotation and ellipticity is improved by using a polarization modulation and lock-in-detection. For the L-MOKE and P-MOKE measurements, the angle of incidence is kept fixed at 70$^{\circ}$ and 45$^{\circ}$, respectively. 

\begin{figure}[ht]
    \centering

    \includegraphics[width=\linewidth]{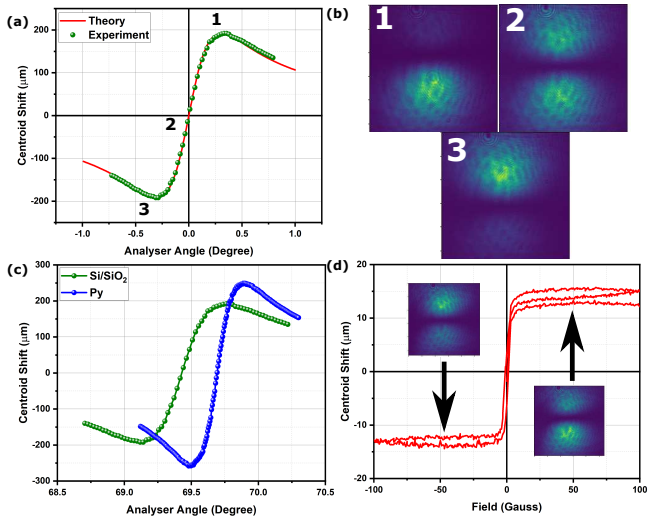}
    \caption{
     (a) A typically measured SHEL shift overlayed with calculated curve is plotted for Si/SiO$_2$ sample with images corresponding to three different analyzer angles. 1 and 3 positions correspond to the maximum intensity of LCP and RCP light, respectively. (b) A comparison of SHEL shift obtained in Py5 and  Si/SiO$_2$ sample indicates a large change in polarization rotation angle. (c) The hysteresis plot was measured while switching 100 Gauss magnetic field. The insets show the two-lobe pattern at maximum shift (right and left polarized components). The intensity of one lobe is higher than the other at the opposite magnetization saturation.}
    \label{fig:fig2}
\end{figure}

\subsection{SHEL-MOKE measurement}
Schematics of the SHEL-MOKE experimental setup is shown in Fig. \ref{fig:sch}(a), where we use a 780 nm diode laser as the source. The laser is passed through a single-mode optical fiber to generate a Gaussian intensity profile beam, with a 1 cm beam diameter. The Gaussian beam is passed through a combination of a long focal length (f = 200 mm) lens (L1) and a Glan-Taylor polarizer (P1) to generate a V-polarized beam with a waist of 16 $\mu$m at focus. The sample placed at the focus reflects the laser beam following the reflection laws explained in section 2. At the air-sample interface, the spin-dependent shift occurs as a result of SOI of light due to the refractive index gradient experienced by the focused beam of light incident at an angle on the sample interface. The reflected beam is then passed through a series of optical elements starting with a quarter wave-plate (QWP) (used in P-MOKE configuration), GT polarizer (P2), and a short focal length (f = 50 mm) lens (L2) to project the spin-dependent shift on the camera. A NIR-sensitive monochromatic CMOS camera is used to measure the position of the image centroid. The measurement of centroid intrinsically involves normalization by the total intensity of light contained in the beam and hence minimizes any intensity fluctuations of the laser. In the setup, the polarizer P1 (called polarizer) and P2 (called analyzer) serve the purpose of pre and post-selection (of the polarization state) of light. The weak amplification is created by keeping their optic axis orthogonal to each other in conjunction with the lens pair used. For the SHEL-MOKE measurement, the sample is placed at the center of the Helmholtz coil, to ensure a uniform field on the surface of the sample. 

The SHEL plot is obtained by measuring the centroid shift of the image as a function of post-selection angle ($\beta$) at zero magnetic field. As mentioned in the previous section,  $\beta_0$ is the value of $\beta$ for which the centroid shift is zero in the SHEL plot. The slope of the linear part of this curve is used as calibration to convert measured centroid shift to Kerr angle (optical rotation).
By fixing the analyzer position at $\beta_0$, the magnetic field is swept between -100 to 100 G. The corresponding change in centroid shift is measured which is a consequence of the change in the SOP of the incident beam due to the Kerr effect. This traces a hysteresis loop, as expected for a ferro-/ferri-magnetic sample. The coercivity of the sample is directly measured from the hysteresis loop. In the current configuration, our system is capable of measuring a minimum optical rotation of 0.03 mdeg. This can be further improved by changing the incident angle, beam waist at the focus, propagation length, etc. The measured magnetic field dependence of the polarization rotation is related to the Kerr rotation and ellipticity induced by the material.

\section{Results and discussion}

Initially, we performed SHEL measurement on Si/SiO$_2$ substrate, which we treat as the reference sample. The Si/SiO$_2$ substrate data is used to confirm our theoretical model developed in section 2. Fig. \ref{fig:fig2}(a) shows the experimentally observed SHEL shift vs change in analyser angle for Si/SiO$_2$ substrate. As can be seen from the plot, the theoretical model agrees well with the experimentally observed variation in the SHEL shift, indicating a strong possibility of characterizing the magnetic and dielectric properties of the sample using this method.

Now, on replacing the Si/SiO$_2$ substrate with substrates coated with ferromagnetic thin film samples, say Py5, the SHEL shift vs analyzer angle changes dramatically as shown in Fig.\ref{fig:fig2}(b) and reveals the high sensitivity of the SHEL shift to accurately characterize the material properties. Since the centroid shift has a linear dependence on the analyzer angle in the region close to the orthogonal analyzer angle with a steep slope, the sensitivity of the system for polarization rotation is highly enhanced. We make use of this to perform sensitive Kerr rotation measurements on our samples. 

\begin{figure}[ht]
    \centering
    \includegraphics[width=\linewidth]{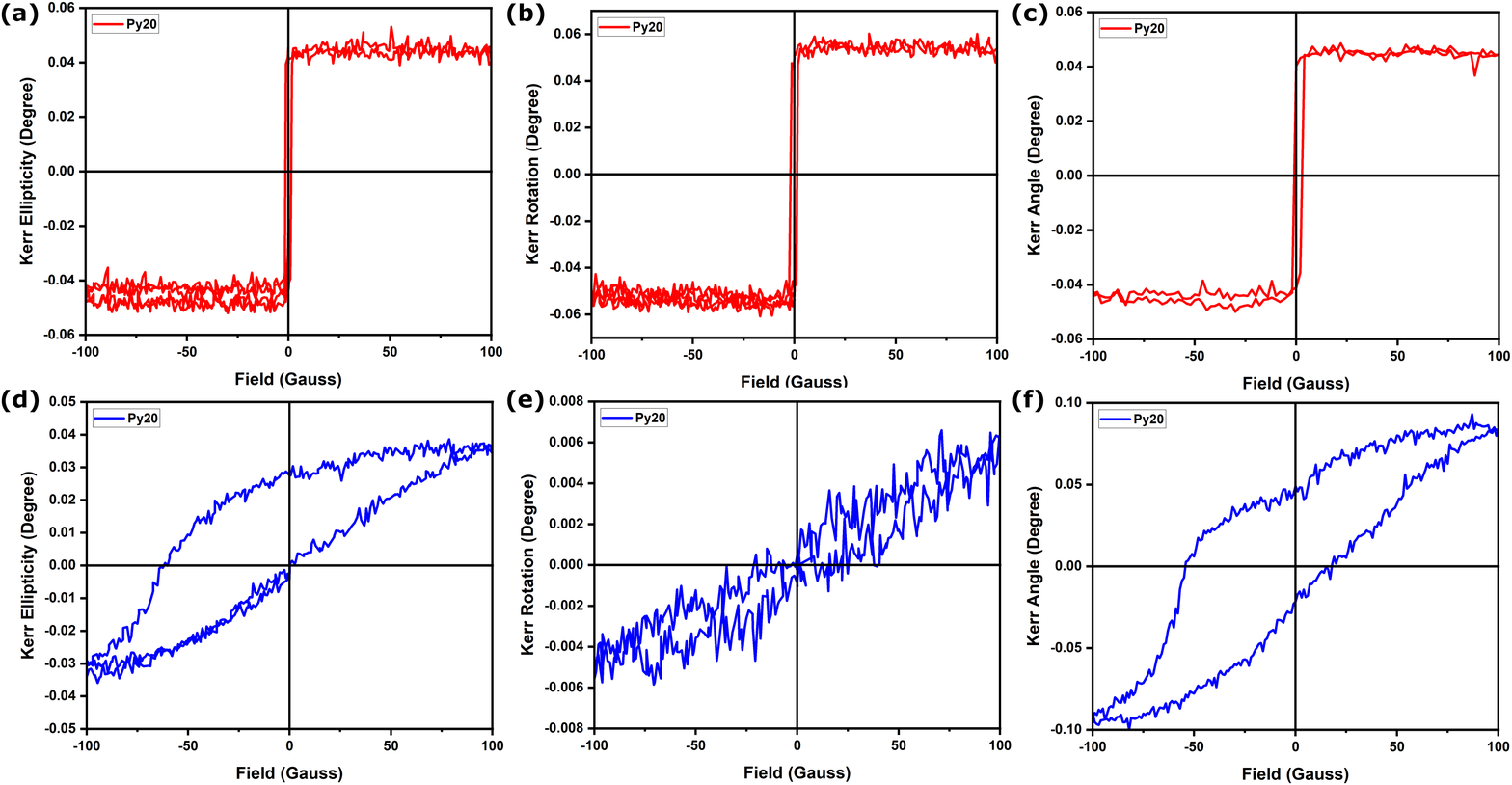}
    \caption{A comparison of MOKE and SHEL-MOKE for  Py20. (a) and (b) show conventional MOKE data for L-MOKE geometry; (d) and (e) for P-MOKE geometry; (c) and (f) show the corresponding SHEL-MOKE data.}
    \label{fig:fig3}
\end{figure} 

To study the magnetic properties of the samples, both L-MOKE and P-MOKE measurements are performed using conventional MOKE and SHEL-MOKE setup. The hysteresis loops are recorded for all the samples by periodically reversing the magnetization direction as shown in Fig.\ref{fig:fig2}(c) and Fig.\ref{fig:fig3} . In Fig.\ref{fig:fig2}(c), keeping analyzer angle at position 2 as denoted in  Fig.\ref{fig:fig2}(a), the changes in centroid shift as a function of applied magnetic field are plotted. On sweeping the magnetic field, the sample magnetization changes and imparts optical polarization rotation on the reflected light, which in turn leads to a centroid shift. The formation of a typical soft magnetic hysteresis loop is evidence from the figure (L-MOKE configuration). The inset shows the intensity profile at two different saturation magnetizations of the Py system. The variation in the intensity of LCP and RCP light at the two extreme magnetic field are evident from Fig.\ref{fig:fig2}(c). 

 The plots in Fig.\ref{fig:fig3}(a) and (b) depict the hysteresis loop of Py20 obtained using conventional L-MOKE experiment, while,  Fig.\ref{fig:fig3}(c) is the hysteresis loop of the corresponding  SHEL-MOKE experiment. In the conventional MOKE experiment, we measure the ellipticity and rotation separately, while the SHEL-MOKE signal depends on both simultaneously. The L-MOKE coercivity is about 1-2 G. In the P-MOKE, the sample shows higher ellipticity change and no hysteresis in the rotation data. The hysteresis in SHEL-MOKE would also be determined by ellipticity change with the magnetic field. To obtain the values of rotation and ellipticity from the SHEL data, we would have to compare the data with theory. The low coercivity seen in L-MOKE data suggests that the easy axis of magnetization lies in the sample plane.

\begin{figure}[ht]
    \centering
    \includegraphics[width=\linewidth]{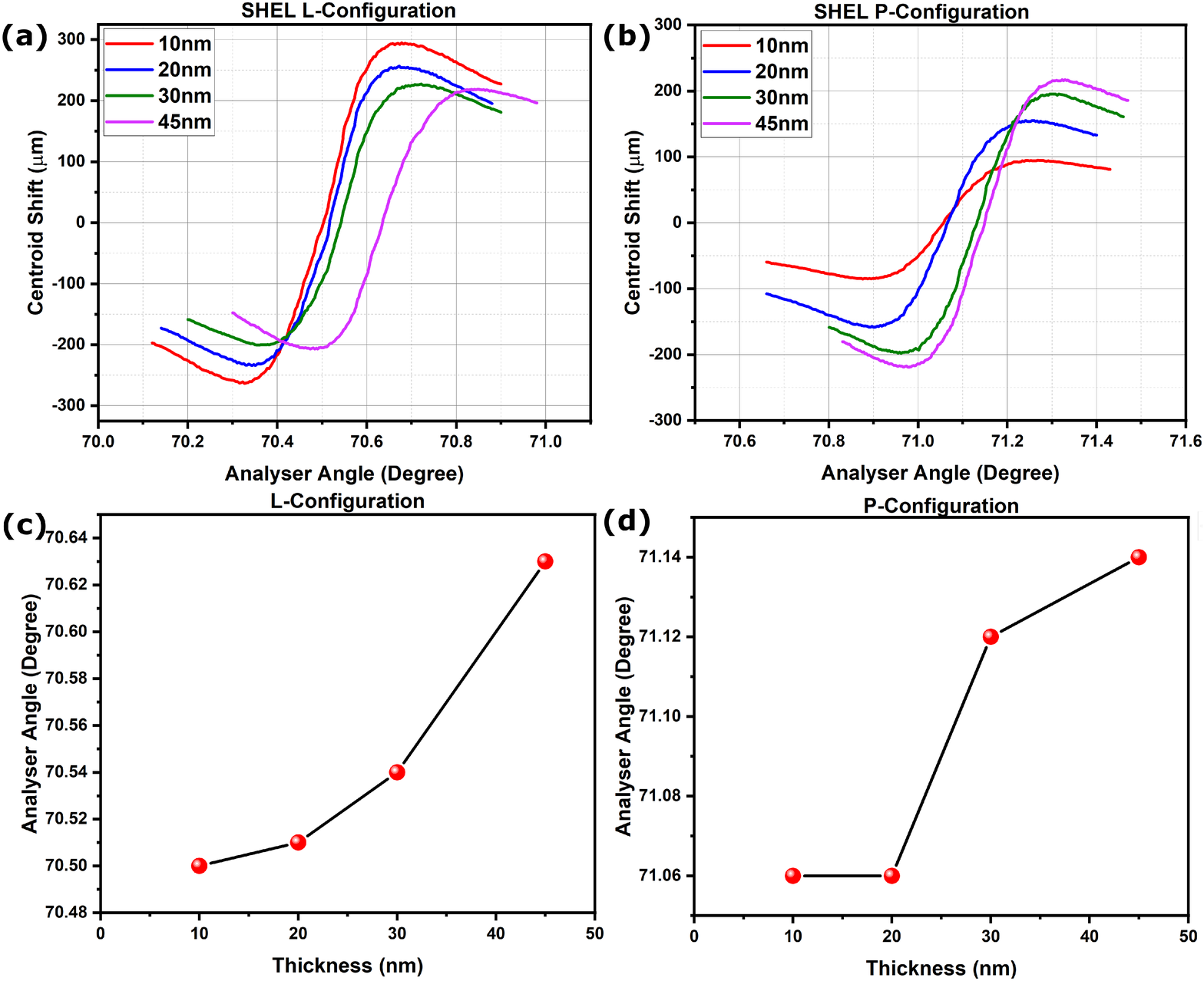}

    \caption{SHEL plot of Py for different thickness in (a) L-configuration and (b) P-configuration. (c) and (d) Change in analyser angle corresponding to zero shift  with thickness in L and P configuration,  respectively. }
    \label{fig:fig4}
\end{figure}

Fig.\ref{fig:fig4}(a) and (b) show the SHEL shift for different thicknesses of Py samples in L-MOKE and P-MOKE configuration, i.e., $\theta_{i}$ is 70$^{\circ}$ and 45$^{\circ}$, respectively at zero magnetic field.  The systematic changes in the centroid shift variation with the thickness of the Py films are evident from the figures. The changes in the analyzer angle corresponding to zero centroid shift with the thickness of films in the two different configurations are shown in Fig.\ref{fig:fig3}(c) and (d), further strengthening the case for the SHEL-based measurement method to accurately and sensitively measure changes in polarization rotation angles in fractions of a degree. Variations in the coercivity with sample thickness observed using SHEL-MOKE are shown in Fig.S4 and Fig.S5 [SI]. The observed coercivity variation in L-MOKE configuration is in the range of 1-3 Gauss for different sample thicknesses. The observance of coercivity in P-MOKE and its variation with sample thickness, as shown in Fig.S6 [SI] proves the soft magnetic nature of Py thin films. 

\begin{figure}[ht]
    \centering
    \includegraphics[width=\linewidth]{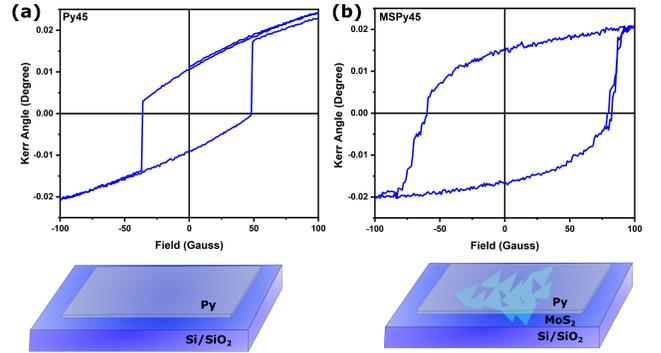}
    \caption{P-MOKE comparison of Py45 and MSPy45. Hysteresis curve of (a) Py45 with 42 G coercivity and (b) MSPy45 where coercivity is around 75 G. Coercivity increases due to SOC of MoS$_2$. Below each figure, the schematics of MSPy heterostructure are provided.}
    \label{fig:fig5}
\end{figure}

Interfacing soft magnetic materials with atomically thin layers to improve the coercivity via spin-orbit coupling is an emerging trend in spintronics due to their fundamental and applied interest \cite{Ahn20202DDevices}. We demonstrate our ability to tune the coercivity of thin films of Py via spin-orbit coupling by interfacing with atomically thin layers of MoS$_2$. The MSPy films developed are now investigated for the changes arising due to SOC by leveraging on the weak amplification-assisted SHEL-MOKE technique. MoS$_2$, being a high SOC material, in proximity to the ferromagnetic materials, modulates the magnetic ordering at the interface due to possible charge/spin transfer\cite{Sahoo2020ProbingMaterials}. Using SHEL-MOKE, we observe this modulation in the magnetic ordering of the permalloy in proximity to MoS$_2$ as a significant enhancement in the observed coercivity in the P-MOKE configuration. A comparison of the coercivity of MSPy samples and Py thin film of similar thickness is shown in Fig.\ref{fig:fig5}. The origin of this enhanced coercivity has been traced to electronic spin and charge transfer between MoS$_2$ and ferromagnets. Hence, magnetic hardening is observed in such an interface.

Furthermore, from the theoretical model, we know that the centroid shift is dependent on the Kerr rotation and Kerr ellipticity introduced by the sample. The zero field SHEL measurement would directly tell us about the rotation and ellipticity contribution to the shift. Permalloy has a complex-valued refractive index, hence the symmetry of the SHEL data in Fig.\ref{fig:fig4} implies that the ellipticity contribution to the shift is negligible. In the case of samples with Kerr ellipticity, the asymmetry observed in the SHEL data would help us estimate the ellipticity contribution. The SHEL data can then be modeled to extract the value of ellipticity and rotation. To study the change in ellipticity as a function of the magnetic field, one would have to measure centroid shift versus analyzer angle at various magnetic fields and evaluate the value of ellipticity from the asymmetry for complex refractive index materials.  Thus SHEL-MOKE can in general be used to measure both Kerr rotation and ellipticity of samples having different dielectric and magnetic properties such as magnetic insulators, non-magnetic insulators, magnetic conductors, non-magnetic conductors, etc.

In addition, the SHEL measurement described above can also be used for sensitive characterisation of the optical conductivity of non-magnetic 2D materials like monolayer MoS$_2$. Existing methods to determine the optical conductivity and susceptibility of MoS$_2$, like the slab approach \cite{slab1,slab2}  or the surface conductivity approach \cite{conductivity1} do not consider the non-planar nature of finite beams and hence cannot give the most accurate charactersation of surface optical properties. Our measurement technique based on SHEL on the other hand, includes the inevitable effect of beam curvature and is a very sensitive technique to determine optical response of MoS$_2$ interfaces for applications in optoelectronic devices.

\begin{figure}[ht]
    \centering
    \includegraphics[width=\linewidth]{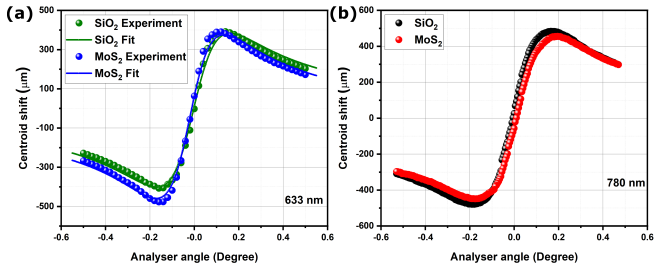}
    \caption{Measured(circles) SHEL shift of monolayer MoS$_2$ mounted on an SiO$_2$ substrate at (a) 633 nm and (b) 780 nm. The shifts at 633 nm show good fit with Eq.\eqref{16}.}
    \label{fig:fig7}
\end{figure}

Fig.\ref{fig:fig7}  (a) and (b) show the observed SHEL shift of SiO$_2$ and monolayer MoS$_2$ grown over SiO$_2$ substrate obtained using 633 nm and 780 nm laser respectively. The SHEL shift for SiO$_2$ with 633 nm are fitted with Eq.\eqref{16} considering Eq.S7, S8 for fresnel reflection coefficients and  shows  a good agreement with the theoritical model. The experimental parameters like the beam waist at focus ($\omega_0$) obtained from fitting of SiO$_2$ SHEL shift is used to obtain the optical conductivity and dielectric sucseptibility of monolayer MoS$_2$ at the given wavelength. From fitting, the optical conductivity of monolayer MoS$_2$ at 633 nm is found to be $(1.9 \cross 10^{-4} \pm 3.3 \cross 10^{-5}) \space \Omega^{-1}$. The obtained value is found in close agreement with the reported value \cite{17,18}.
From the figure, we note that the ellipticity introduced by MoS${_2}$ monolayer, given by the change in slope of SHEL shift is greater than that of bare SiO$_2$ at 633 nm and lesser than SiO$_2$ at 780 nm. This wavelength dependent ellipticity variation is an unexplored field of the SHEL technique, both theoretically and experimentally, and will be explored in the near future.

Thus, the veracity of the theoretical model is established and it is shown that it can be a powerful method for analysing ultra-thin samples' magnetic, dielectric, and optical properties with a high spatial resolution. 

\section{Conclusion}
With a detailed theoretical analysis and simulations provided here, we establish the potential of SHEL-MOKE as a high-sensitivity simple surface probing tool for studying materials' magnetic and dielectric properties. We experimentally demonstrated the versatility of the SHEL-MOKE method by characterizing the surface and interface magnetic properties of ultra-thin film samples of permalloy (Py) and MoS$_2$-Py interfaced samples. The ability to characterize the real and imaginary parts of the complex Kerr angle is expected to further improve the applicability of the method, and enable researchers to characterize accurately the effects arising at the magnetic-nonmagnetic interface. The coercivity measurement done using the SHEL-MOKE provides a direct and non-perturbative measurement technique to probe SOC-induced effects in material interfaces. Variations in the surface-interface conductivity, anticipated by the theoretical calculations, are currently being investigated, which adds value to the proposed methodology. The capability to extract surface-interface magnetization, coercivity, and conductivity from a single SHEL-MOKE scan with significantly improved sensitivity combined with simplicity will no doubt make the method a potential successor to several existing and widely used methods. At the end, the method has been extended to other ultra-thin films such as monolayer MoS$_2$, where its optical conductivity is extracted, indicating the potential of this method in analysing thin films of different kinds.

\section{acknowledgement}
 Authors from TIFR thank the support of the Department of Atomic Energy, Government of India, under Project Identification No. RTI 4007. GR. TNN and NKV thank SERB-SUPRA for suppport under grant no. SPR/2020/000220. NKV thanks SERB-CRG  and IOE-UOH for financial support to this area of research. JJP acknowledges Sushree S. Sahoo and Parswa Nath for useful discussions.

\section{Supplementary Information}
 See Supplementary Information [SI] for supporting  details about theory,  material characterization, and additional data.

\section{Data Availability}
The data underlying this study are available in the published article and its online supplementary material.

\bibliography{references}

\end{document}


\title{High-sensitivity characterization of ultra-thin magnetic samples using spin-Hall effect of light: supplementary material}

\author[1]{Janmey J. Panda}

\author[1]{Krishna R. Sahoo}

\author[1]{Aparna Praturi}

\author[1]{Ashique Lal}

\author[2]{Nirmal K. Viswanathan}


\author[1]{Tharangattu N. Narayanan}

\author[1]{G. Rajalakshmi}
\affil[1]{Tata Institute of Fundamental Research - Hyderabad, Sy. No 36/P Serilingampally Mandal, Gopanpally Village, Hyderabad 500046, India}
\affil[2]{School of Physics, University of Hyderabad, Hyderabad 500046, Telangana, India.}


\date{}
\maketitle

\section{SHEL Modified Shift for Magnetic material}
For V-polarized incident light the dependence of complex zero field rotation angle $\theta_{V}$ on modified shift for a gyrotopic media is given by:

\begin{equation}
\begin{split}
&\Delta_{V}^{mod}=\\&\frac{-2z(\delta_{V}^{\prime}\theta_{V}+\delta_{V}^{\prime *}\theta_{V}^{*})-ik(\delta_{V}^{\prime}\theta_{V}-\delta_{V}^{\prime *}\theta_{V}^{*})\omega^{2}+(2z(\delta_{V}^{*}\theta_{V}+\delta_{V}\theta_{V}^{*})+ik(-\delta_{V}^{*}\theta_{V}+\delta_{V}\theta_{V}^{*})\omega^{2})\cot^{2}{\beta}}
{2k(\zeta_{V}^{2}+\delta_{V}^{2}\theta_{V}^{2}+\omega^{2}+((\zeta_{V}^{*}\zeta_{V}^{\prime}-\delta_{V}^{*}\delta_{V}^{\prime})\theta_{V}+(\zeta_{V}\zeta_{V}^{\prime *}-\delta_{V}\delta_{V}^{\prime *})\theta_{V}^{*}+(\theta_{V}+\theta_{V}^{*})\omega^{2})\cot{\beta})+(\delta_{V}^{2}+\theta_{V}^{2}(\omega^{2}+\zeta^{\prime 2}_{V}))\cot^{2}{\beta}))}
\\
&+\frac{(2z(\delta_{V}+\delta_{V}^{*}-(\delta_{V}^{\prime}+\delta_{V}^{\prime *})\theta_{V}^{2})+ik(\delta_{V}-\delta_{V}^{*}+(-\delta_{V}^{\prime}+\delta_{V}^{\prime *})\theta_{V}^{2})\omega^{2})\cot{\beta}}
{2k(\zeta_{V}^{2}+\delta_{V}^{2}\theta_{V}^{2}+\omega^{2}+((\zeta_{V}^{*}\zeta_{V}^{\prime}-\delta_{V}^{*}\delta_{V}^{\prime})\theta_{V}+(\zeta_{V}\zeta_{V}^{\prime *}-\delta_{V}\delta_{V}^{\prime *})\theta_{V}^{*}+(\theta_{V}+\theta_{V}^{*})\omega^{2})\cot{\beta})+(\delta_{V}^{2}+\theta_{V}^{2}(\omega^{2}+\zeta^{\prime 2}_{V}))\cot^{2}{\beta}))}
\\
\end{split}.
\end{equation}

For H polarized incident light the dependence of complex zero field rotation angle $\theta_{H}$ on modified shift for a gyrotopic media is given by:

\begin{equation}
\begin{split}
&\Delta_{H}^{mod}=\\&\frac{-2z(\delta_{H}^{\prime}\theta_{H}+\delta_{H}^{\prime *}\theta_{H}^{*})+ik(\delta_{H}^{\prime}\theta_{H}-\delta_{H}^{\prime *}\theta_{H}^{*})\omega^{2}-(2z(\delta_{H}^{*}\theta_{H}+\delta_{H}\theta_{H}^{*})+ik(-\delta_{H}^{*}\theta_{H}+\delta_{H}\theta_{H}^{*})\omega^{2})\cot^{2}{\beta}}{2k(\zeta_{H}^{2}+i(\zeta^{\prime *}_{H}\zeta_{H}\theta_{H}^{*}-\zeta^{\prime}_{H}\zeta^{*}_{H}\theta_{H})+(\delta_{H}^{2}+\zeta^{\prime 2}_{H})\theta_{H}^{2}+\omega^{2}+(\delta_{H}^{*}\delta_{H}^{\prime}\theta_{H}+\delta_{H}\delta_{H}^{\prime *}\theta_{H}^{*}-(\theta_{H}+\theta_{H}^{*})\omega^{2})\cot{\beta})+(\delta_{H}^{2}+\theta_{H}^{2}\omega^{2})\cot^{2}{\beta}))}
\\+&\frac{(2z(\delta_{H}+\delta_{H}^{*}-(\delta_{H}^{\prime}+\delta_{H}^{\prime *})\theta_{H}^{2})+ik(\delta_{H}-\delta_{H}^{*}+(-\delta_{H}^{\prime}+\delta_{H}^{\prime *})\theta_{H}^{2})\omega^{2})\cot{\beta}}{2k(\zeta_{H}^{2}+i(\zeta^{\prime *}_{H}\zeta_{H}\theta_{H}^{*}-\zeta^{\prime}_{H}\zeta^{*}_{H}\theta_{H})+(\delta_{H}^{2}+\zeta^{\prime 2}_{H})\theta_{H}^{2}+\omega^{2}+(\delta_{H}^{*}\delta_{H}^{\prime}\theta_{H}+\delta_{H}\delta_{H}^{\prime *}\theta_{H}^{*}-(\theta_{H}+\theta_{H}^{*})\omega^{2})\cot{\beta})+(\delta_{H}^{2}+\theta_{H}^{2}\omega^{2})\cot^{2}{\beta}))}
\end{split}.
\end{equation}
Here, real and imaginary part of $\theta_{V}$ and $\theta_{H}$ represent rotation and ellipticity respectively.
\newpage
\section{Fresnel reflection coefficients for magneto optic materials}
For a magneto-optical material the fresnel coefficient can be written as: 

\begin{equation}
    r^{pp}=\frac{n_2 \cos{\theta_i}- n_1 \cos{\theta_t}+i n_1 Q_{y}\sin{\theta_t}}{n_1 \cos{\theta_t}+ n_2 \cos{\theta_i}-i n_1 Q_{y}\sin{\theta_t}}
\end{equation}

\begin{equation}
    r^{ss}=\frac{n_1 \cos{\theta_i}- n_2 \cos{\theta_t}}{n_1 \cos{\theta_i}+ n_2 \cos{\theta_t}}
\end{equation}

\begin{equation}
    r^{sp}=-\frac{i n_1 n_2\cos{\theta_i}\sec{\theta_t}(Q_z \cos{\theta_t}+Q_x \sin{\theta_t})}{(n_1 \cos{\theta_i}+ n_2 \cos{\theta_t})(n_1 \cos{\theta_t}+ n_2 \cos{\theta_i}-i n_1 Q_{y}\sin{\theta_t})}
\end{equation}

\begin{equation}
    r^{ps}=\frac{i n_1 n_2\cos{\theta_i}\sec{\theta_t}(-Q_z \cos{\theta_t}+Q_x \sin{\theta_t})}{(n_1 \cos{\theta_i}+ n_2 \cos{\theta_t})(n_1 \cos{\theta_t}+ n_2 \cos{\theta_i}-i n_1 Q_{y}\sin{\theta_t})}
\end{equation}
\section{Fresnel reflection coefficients for non-magnetic ultrathin dielectric layer}
For a ultrathin layer of a dielectric material with susceptibility $\chi$  and optical conductivity $\sigma$ deposited on a substrate of refractive index $n_2$, the fresnel coefficients can be written as: 

\begin{equation}
    r^{ss}=\frac{n_1 \cos{\theta_i}- n_2 \cos{\theta_t}-i k\chi-\sigma Z_0}{n_1 \cos{\theta_i}+ n_2 \cos{\theta_t}+i k\chi+\sigma Z_0}
\end{equation}

\begin{equation}
    r^{pp}=\frac{n_2 \cos{\theta_i}- n_1 \cos{\theta_t}+(i k\chi+\sigma Z_0)cos{\theta_i}cos{\theta_t}}{n_2 \cos{\theta_i}+ n_1 \cos{\theta_t}+(i k\chi+\sigma Z_0)cos{\theta_i}cos{\theta_t}}
\end{equation}	1

where $Z_0$ denotes the impedance of vacuum.
\newpage
\section{Sample characterization}
\subsection{SEM and EDS characterization of Py thin film}
\begin{figure}[htbp]
\centering
\includegraphics[width=.95\linewidth]{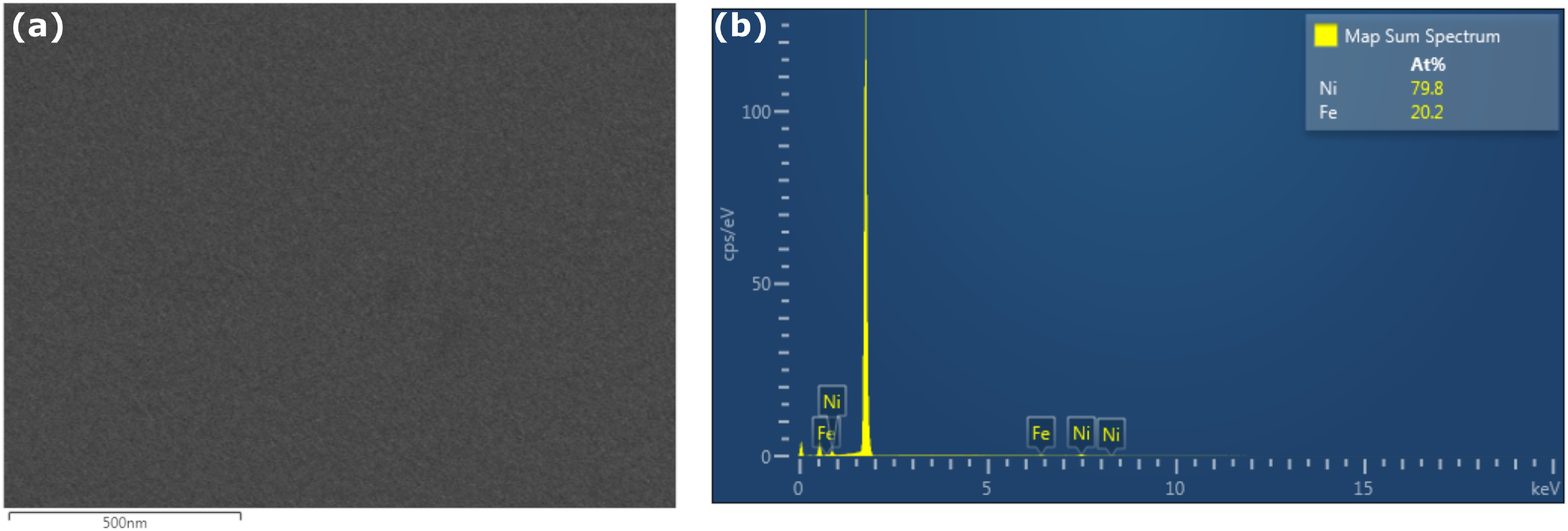}
\caption{(a) SEM image of the film. (b) Atomic percentage from SEM-EDS analysis showing Ni $\sim$ 80\% and Fe $\sim$ 20 \% in Py samples.}
\label{fig:sem}
\end{figure}
\subsection{Raman Spectroscopy of MoS$_{2}$(MS) and MSPy sample}
\begin{figure}[h]
\centering
\includegraphics[width=.50\linewidth]{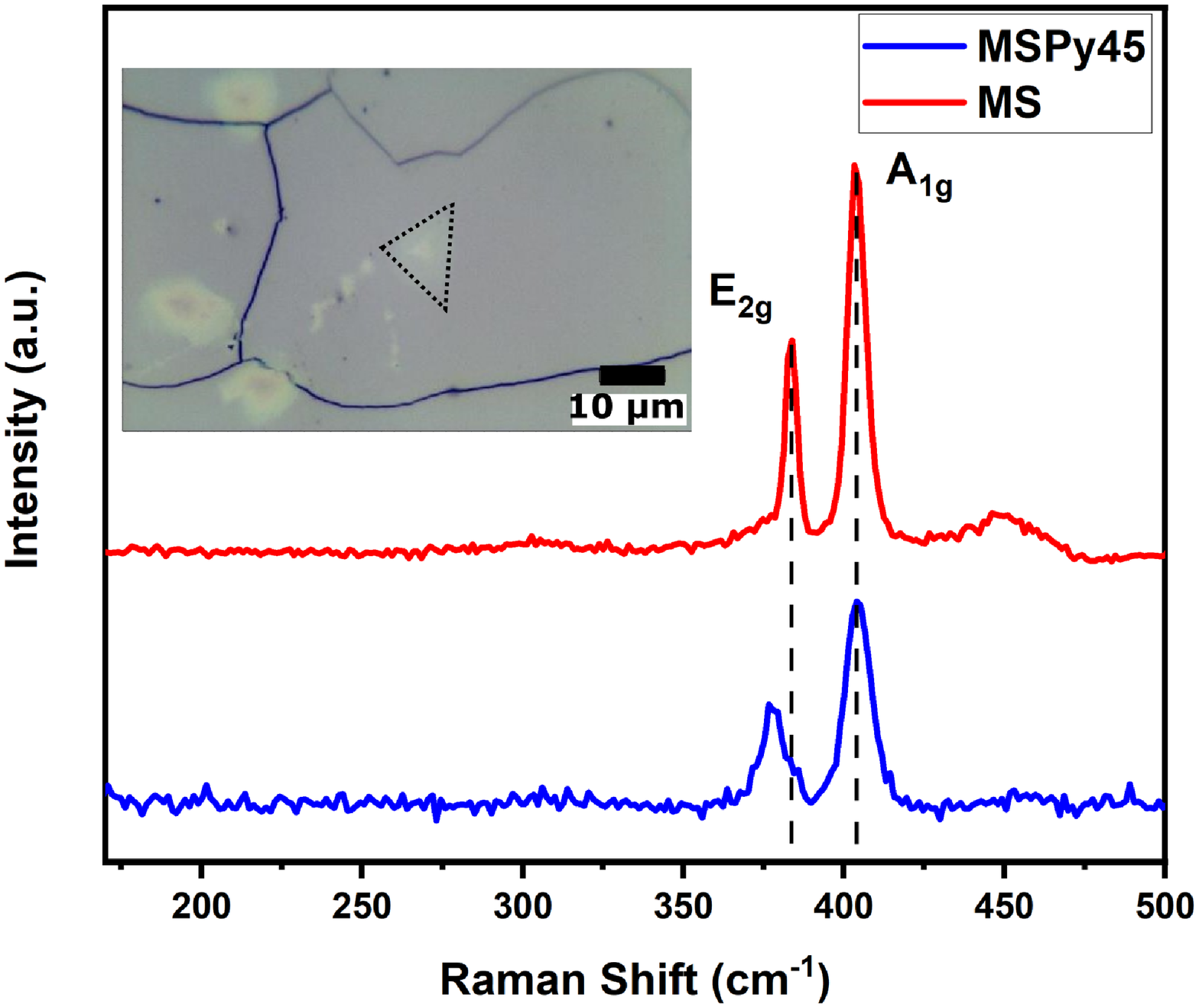}
\caption{Raman spectra of MS and MSPy45. The peak position of A$_{1g}$ is same where as there is a red shift in E$_{2g}$ peak position because of strain induced on MoS2 due to permalloy deposition. The insert shows the optical image of MSPy45. Absence of extra peak confirms that MoS$_2$ was retaining in its 2H phase after Py  deposition. }
\label{fig:raman}
\end{figure}

\newpage

\subsection{Conventional MOKE setup}
\begin{figure}[htbp]
\centering
\includegraphics[width=.95\linewidth]{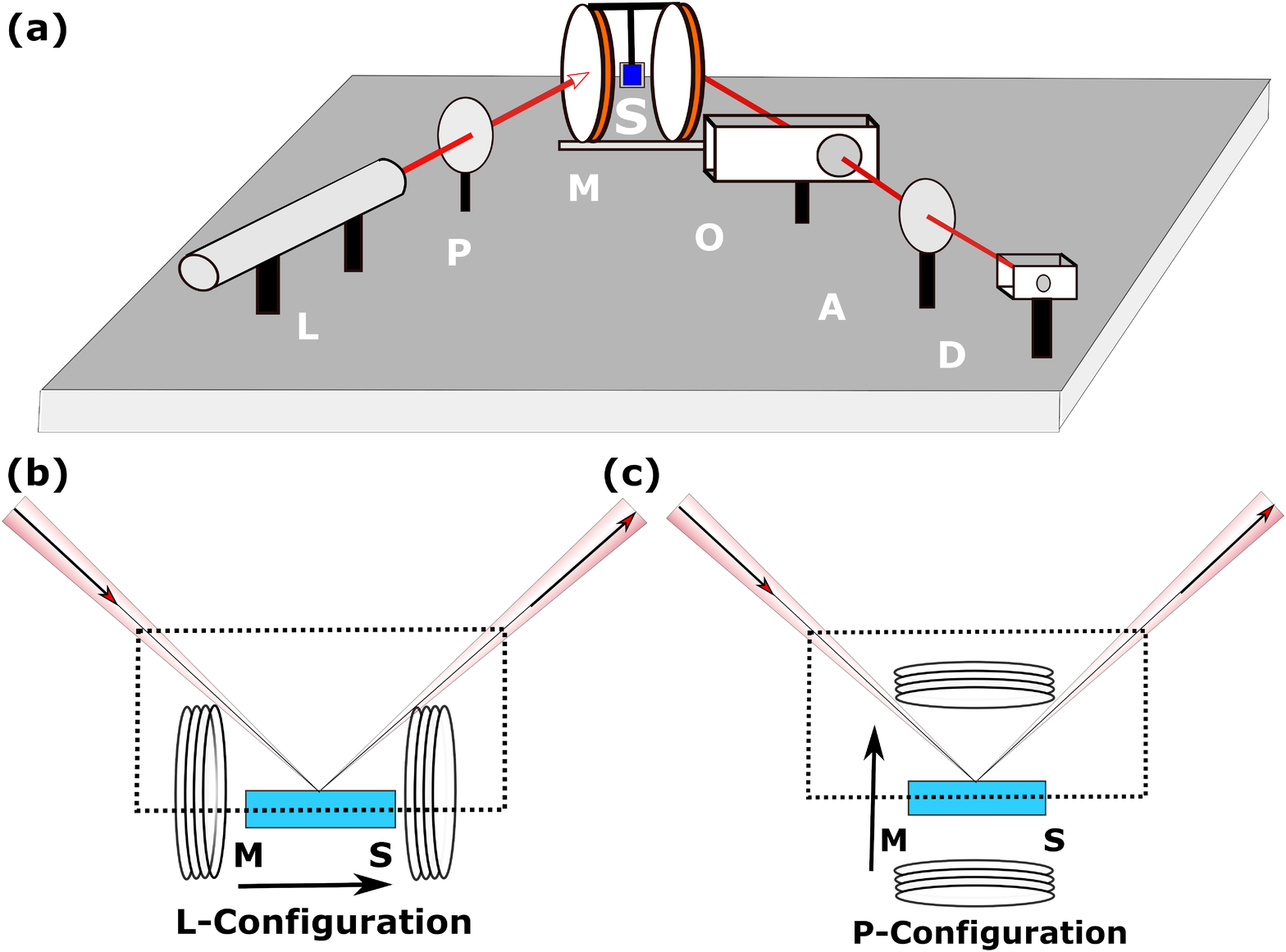}
\caption{(a) Schematic of conventional MOKE setup. L- Laser source (633 nm), P- Polariser, M- Magnet, S- Sample, O- Modulator, A- Analyser, D- Detector.(b) Longitudinal MOKE (L-MOKE) geometry. M is magnetisation direction parallel to sample. (c) Polar MOKE  (P-MOKE) geometry. M is magnetisation direction perpendicular to sample.  }
\label{fig:moke}
\end{figure}

\newpage

\subsection{ Magnetization hysteresis obtained using SHEL-MOKE}
\begin{figure}[htbp]
\centering
\includegraphics[width=.85\linewidth]{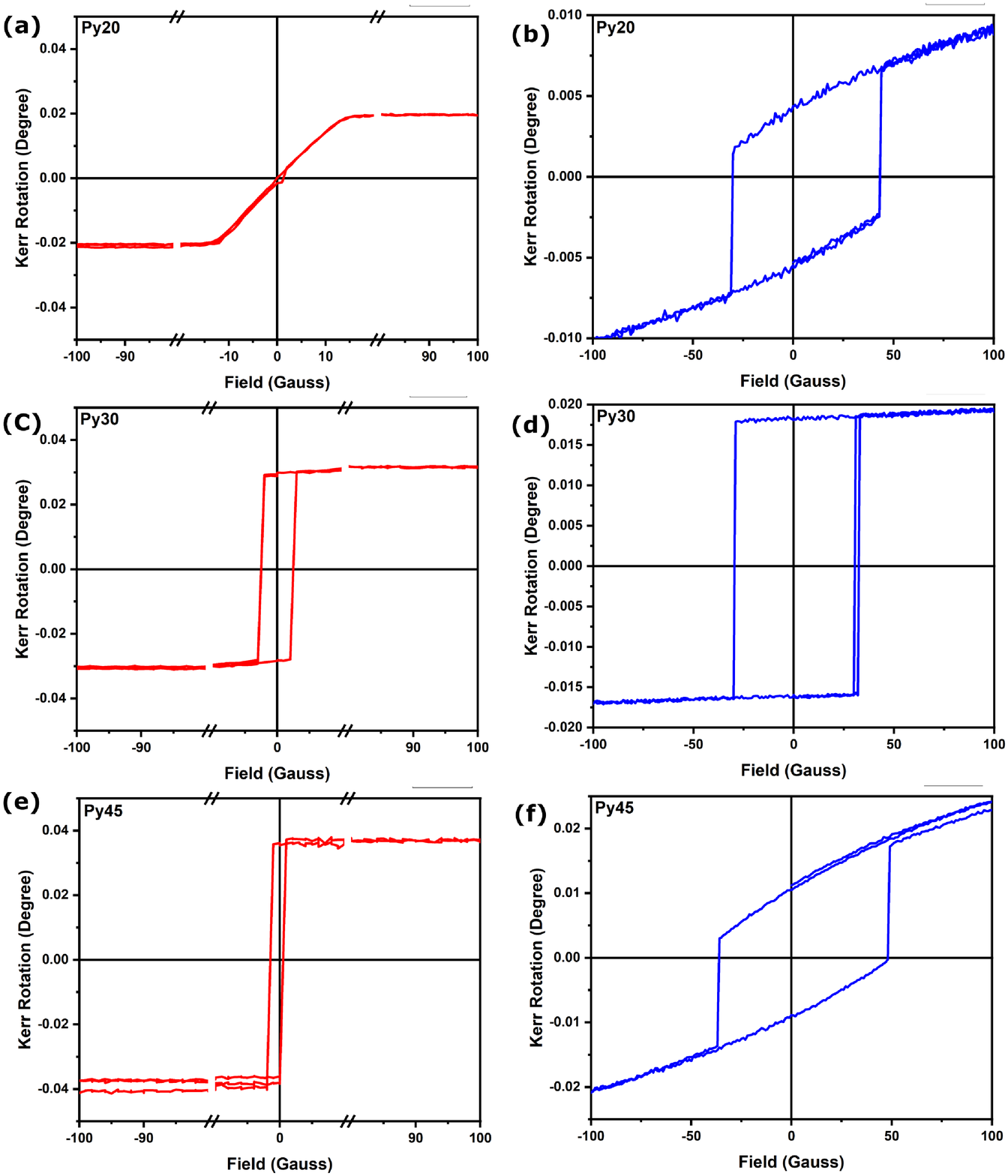}
\caption{ SHEL-MOKE data for different thicknesses of Py in L-configuration (a, c, e) and P-configuration   (b, d, f).}
\label{fig:lmoke}
\end{figure}

\newpage
\subsection{Variation of coercivity with thickness of the permalloy thin film }
\begin{figure}[htbp]
\centering
\includegraphics[width=.85\linewidth]{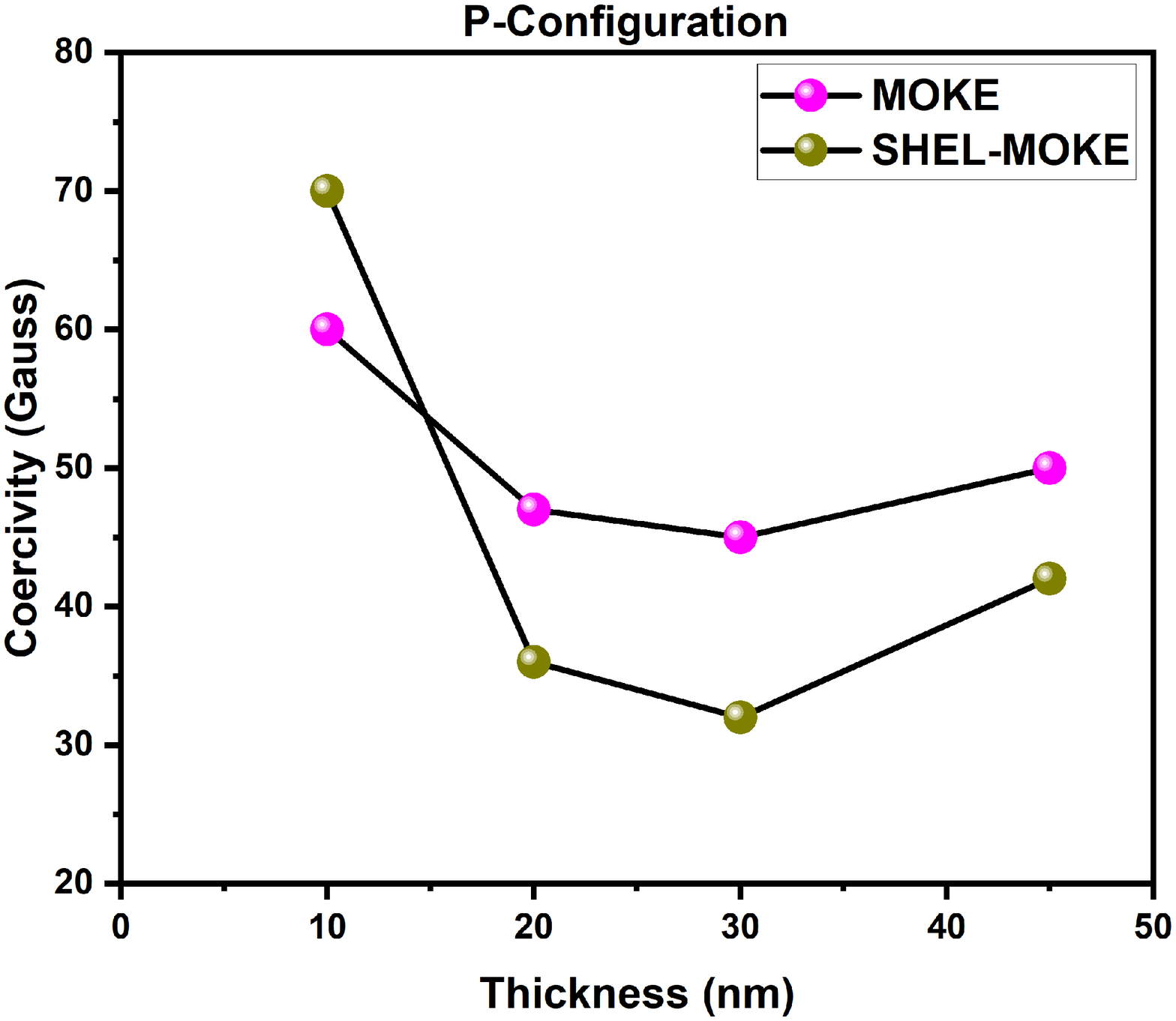}
\caption{Coercivity vs. thickness plot for Py10, Py20, Py30, Py45  in P-configuration. In both MOKE and SHEL-MOKE show similar types of behaviour. The MOKE data shows ellipticity measurement.}
\label{fig:coer}
\end{figure}
\newpage
\subsection{Variation in SHEL Slope and centroid shift range with different thickness of permalloy}
\begin{figure}[htbp]
\centering
\includegraphics[width=.95\linewidth]{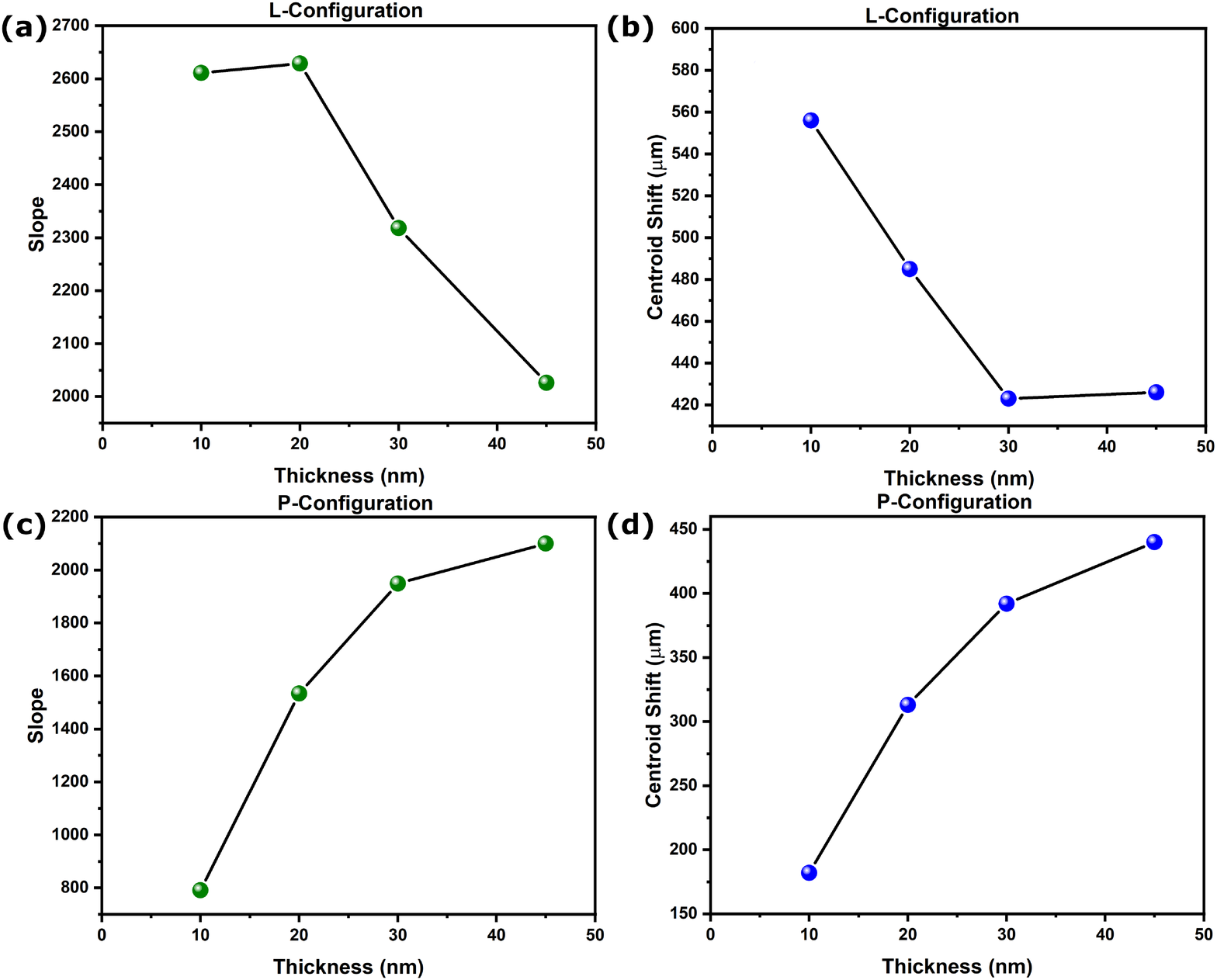}
\caption{Slope changes and centroid shift range are plotted for different thicknesses of Py in (a-b) L-configuration, (c-d) P-configuration.
}
\label{fig:slope}
\end{figure}

    The changes in the slope of the centroid shift variation with analyzer angle position for different thickness of Py are plotted in \ref{fig:slope}~(a-b). The difference between the maximum and minimum centroid shift measured for each sample corresponding to position 1, 3 in Fig.2~(b) of main text are  also plotted for different thickness in  \ref{fig:slope}~(c-d)
